\documentclass[traditabstract]{aa}
\usepackage{txfonts}
\usepackage{graphicx}

\newcommand\phs{\phantom{$-$}}
\newcommand{\rb}[1]{\raisebox{1.5ex}[-1.5ex]{#1}}

\newcommand{\wcen}{$\omega$Cen~}
\newcommand{\wcenp}{$\omega$Cen.~}

\begin{document}

\title{Purveyors of fine halos} 
\subtitle{III. Chemical abundance analysis of a potential \wcen associate\thanks{Based on observations obtained at ESO Paranal Observatory, program 0104.D-0059.}}

\author{
Andreas J. Koch-Hansen\inst{1},  
Camilla Juul Hansen\inst{2},
Linda Lombardo\inst{3},
Piercarlo Bonifacio\inst{3},
Michael Hanke\inst{1},
Elisabetta Caffau\inst{3}
}
\authorrunning{A.J. Koch-Hansen et al.}
\titlerunning{Abundance analysis of an \wcen associate}
\offprints{A.J. Koch-Hansen;  \email{andreas.koch@uni-heidelberg.de}}
\institute{
Zentrum f\"ur Astronomie der Universit\"at Heidelberg, Astronomisches Rechen-Institut, M\"onchhofstr. 12, 69120 Heidelberg, Germany 
\and Max-Planck Institut f\"ur Astronomie, K\"onigstuhl 17, 69117 Heidelberg, Germany
\and GEPI, Observatoire de Paris, Universit\'e PSL, CNRS, 5 Place Jules Janssen, 92190, Meudon, France
}
\date{}
\abstract{
Globular clusters (GCs) are important donors to the build-up of the Milky Way (MW) stellar halo, having contributed at the ten percent level over 
the Galactic history. 
Stars that originated from the second generation of
dissolved or dissolving clusters can be readily identified via distinct light-element signatures such as enhanced
N and Na and simultaneously depleted C and O abundances.
In this paper we present an extensive chemical abundance analysis of 
the halo star J110842, which was previously   kinematically associated with the massive MW GC $\omega$ Centauri ($\omega$Cen), and 
we discuss viable scenarios from escape to encounter.
Based on a high-resolution, high signal-to-noise  spectrum of this star using the UVES spectrograph, we were able to  
measure 33 species of 31 elements across all nucleosynthetic channels. 
The star's low metallicity of [Fe\,{\sc ii}/H]=$-2.10$$\pm$0.02(stat.)$\pm0.07$(sys.) dex places it in the lower sixth percentile of 
$\omega$Cen's metallicity distribution. We find that all of the heavier-element abundances, from $\alpha$- and Fe-peak elements to 
neutron-capture elements are closely compatible with $\omega$Cen's broad abundance distribution. However, given the major overlap of
this object's abundances with the bulk of all of the MW components, this does not allow for a clear-cut distinction of the star's origin.
In contrast, our measurements of an enhancement in CN and its position on the Na-strong locus of the Na-O anticorrelation render it 
conceivable that it   originally formed as a second-generation GC star, lending support to a former association of this halo star with the massive GC \wcenp
 }
\maketitle
%
%
%
%
\section{Introduction}
The stellar halo of the Milky Way (MW) galaxy  conceivably formed through a variety of channels.
Thus, {in situ} star formation within the host galaxy is contrasted by an {ex situ} formation, where the 
halo stars were born in satellite galaxies and accreted onto the host system only later on.
The purported relative importance of either scenario varies in the literature and it is currently believed that our 
Galaxy experienced a mixture of both, where the {ex situ} component contributed to different degrees
depending on galactocentric radius \citep{Eggen1962,SearleZinn1978,Dekel1986,Bullock2005,Zolotov2009,Cooper2013,Pillepich2015,Naidu2020}.

One important class of donors to the buildup
 of the MW halo is the globular clusters (GCs), and there is a wealth of 
evidence  for their ongoing tidal disruption and for their accretion, ranging from observations of stellar streams 
\citep[e.g.,][]{Odenkirchen2001,Lee2004} and extended envelopes of present-day GCs \citep[e.g.,][]{Jordi2010,Kuzma2018}
to the chemodynamical identification of former GC stars in the MW halo field \citep{Martell2010,Koch2019CN,FernandezTrincado2019,Tang2019,Hanke2020} 

The key signature to identify a  {bona fide} cluster escapee lies in the chemical anomalies inherent in the multiple populations of the GSs  \citep{Carretta2009NaO,Milone2017,Bastian2018,Gratton2019}.
As a result of high-temperature proton-capture reactions in the CNO cycle and its Ne-Na chain in a first generation of massive stars, 
the second generation that forms from the ejecta of these polluters is found to be rich in He, N, Na, and Al, while depleted in C, O, and Mg, which 
leads to the characteristic anticorrelations (Na-O, Mg-Al) and bimodalities (e.g., in CN) observed in any given GC \citep{Cohen1978,Carretta2009NaO,Hanke2017,Bastian2018}.
The remainder of the chemical inventory of the second stellar generation remains largely unaltered by the involved nuclear reactions. These chemical patterns are indeed
the best tracers of GC escapees,
 provided they were part of the second generation, 
as these characteristic abundances
are predominantly found in GCs across the entire mass range, while absent in young open clusters, dwarf galaxies, and 
{in situ} halo field stars \citep{Pilachowski1996,Geisler2007,Bragaglia2017,Bekki2019}.

Based on the these chemical signatures, 
recent quantitative analyses
have estimated that about 11\% of the stellar MW halo   originated from now defunct GCs \citep{Martell2010,Martell2011,Koch2019CN,Hanke2020}\footnote{ 
This value depends, among other factors, on the adopted fraction of 
first-generation stars that were lost from the GCs at early times. While \citet{Koch2019CN} adopt a fraction of 56\%, the recent analysis of \citet{Hanke2020} 
assigns larger values of 50--80\%, which would raise the inferred halo fraction to the 20\% level.}, 
an order of magnitude that is bolstered by simulations \citep[e.g.,][]{ReinaCampos2020}. 
Typically, such studies employ low-resolution spectroscopy, which is suitable to determine CN-band strengths, but is not sufficient to 
perform detailed chemical abundance analyses that inform us about the chemical properties of the progenitor cluster \citep[cf.][]{Ramirez2012,Lind2015,Hendricks2016,Majewski2017,Fernandez-Trincado2016,Fernandez-Trincado2017}. 
Taking the chemical identification of potentially former GC stars in the halo field one step further, \citet{Hanke2020} added the kinematic dimensions 
afforded by the astrometry from the second data release (DR2) of the {\em Gaia} mission 
\citep{Gaia2018Lindegren,GaiaDR2}. This allowed us not only to detect extra-tidal stars around known GCs, but also to 
trace back stars with common phase-space portions to potentially common progenitors \citep[see also][]{Savino2019}. 
In addition to disrupting GCs in the present-day MW halo, we also need to consider the accretion of (dwarf) galaxy satellites. On the one hand, 
this leads to the donation of their GC systems, thereby increasing the census of the GC population in the Galaxy \citep{Cohen2004,Forbes2010,Law2010,Carretta2017,Massari2019,Myeong2019,Kruijssen2019,Koch2019Pal13,Forbes2020}.
On the other hand, the most massive GCs in the MW system are often considered the former nuclei of disrupted dwarf galaxies, leading to 
broad abundance spreads and pronounced multiple populations that are unequalled in the lower-mass star clusters
\citep{Bassino1995,Sarajedini1995,Forbes2004,Hilker2004,Kayser2006,Johnson2010}. Of these clusters 
the most massive GC in the MW system, $\omega$~Centauri ($\equiv$ NGC 5139; herafter $\omega$Cen), has long   been discussed as the core of a dwarf galaxy satellite
\citep[e.g.,][]{Lee1999,Bekki2003,Romano2007,Valcarce2011}; its metallicities  show a broad range  from $-2.5$ to $-0.8$ dex and its chemical abundance inhomogeneities 
are many \citep{Freeman1975,Cohen1981,Suntzeff1996,Hilker2004,McWilliam2005,Johnson2010,Villanova2014,Magurno2019,Johnson2020wcen}. 

Based on a large spectroscopic sample from the RAdial Velocity Experiment (RAVE; \citealt{Steinmetz2006,Kordopatis2013}), 
\citet{FernandezTrincado2015} kinematically associated 15 halo stars with \wcen that were either subject to high-velocity ejections from it some 200~Myr ago or 
that had had close encounters with this particular GC  at high relative velocities. 
In this work we present a high-resolution, high signal-to-noise (S/N) chemical abundance analysis of one of these candidates, the metal-poor ([Fe/H]$\sim -2$ dex)  field
star J110842.1$-$715260 (hereafter J110842).\footnote{This star was observed as part of the CERES project (``Chemical Evolution of R-process Elements in Stars'') and thus has the alternate ID CES1108$-$7153 (C.J. Hansen et al., in prep.).}  
As the relative velocity in the close encounter with $\omega$Cen, at $v_{\rm rel}$=275 km\,s$^{-1}$, is rather high,  
\citet{FernandezTrincado2015} concluded that it is unlikely that this star (and other similar ones)   directly escaped from the GC,  
rather that it encountered and interacted with the GC between 45 and 290~Myr ago\footnote{Depending on the adopted Galactic potential.
The orbital analysis of \citet{FernandezTrincado2015} also relied on proper motions from the UCAC4 \citep{Zacharias2013}, which, however, are in good agreement 
with the latest {\em Gaia} values.}. 
The full sample of the RAVE-$\omega$Cen associates shows chemical abundances that are consistent with the stellar populations of this object; 
however, 
the RAVE spectra only allowed   the determination of the $\alpha$-elements Mg, Si, and Ti, and Al and Ni. For J110842 only Al and Ni could be determined. 
Therefore, a more complete sampling of the abundance space has yet to be conducted, in order to investigate whether the 
origin of this star  is indeed similar to that of $\omega$Cen
stars and, if so, whether it classifies as a first- or second-generation star.

This paper is organized as follows. In Sect.~2 we describe the data, followed by details of the abundance analysis in Sect.~3. The resulting chemical abundances are presented in 
Sect.~4. We discuss the results in the context of  $\omega$Cen's chemodynamical properties in Sect.~5. 
\section{Data}
Star J110842 was observed on the night of March 03, 2020, with the Ultraviolet and Visual Echelle Spectrograph
(UVES; \citealt{Dekker2000}) at the Very Large Telescope (Program 0104.D-0059; P.I. C.J. Hansen). 
We employed the 390/564 setting with dichroic\,1, leading to a broad wavelength coverage of 
3285--4518, 4623--5600, and 5672--6647 \AA, and a high spectral resolving power of $R\sim40000$.
An exposure time of 2700 seconds 
yielded a S/N of $\sim$50, 120, and 150 per pixel at 4000, 5200, and 6200 \AA.
The data were reduced in the standard manner 
using the ESO UVES reduction workflow recipes (version 6.1.3) that perform 
bias correction, order tracing, flat fielding, and wavelength calibration using calibration data 
that were obtained on the same day as the observations.

We measured the star's radial velocity from a cross-correlation against a template of similar stellar
parameters  using the  Image Reduction and Analysis Facility (IRAF) {\em fxcor} task. 
This yielded a heliocentric velocity of  $v_{\rm HC}$=273.1$\pm$0.1 km\,s$^{-1}$, which agrees well with 
the values reported from the lower-resolution RAVE and {\em Gaia} spectra to within 0.4 km\,s$^{-1}$ (see Table~1). 
\begin{table}[htb!]
\caption{Properties of the target star.}
\centering
\begin{tabular}{ccc}
\hline\hline       
Parameter & Value & Reference \\
\hline                  
$\alpha$ (J2000.0) & \phs11:08:42.12 & 1 \\
$\delta$ (J2000.0) & $-$71:52:59.9 & 1 \\
$G$ & 11.117 & 2 \\
$G_{BP}$ & 11.117 & 2 \\
$G_{RP}$ & 10.239 & 2 \\
$L$ & 1224 $L_{\odot}$ & 2 \\
$v_{\rm HC}$ & 273.1$\pm$0.1 km\,s$^{-1}$ & 3 \\
$T_{\rm eff}$ & 4421$\pm$50 K& 3 \\
log\,$g$  & 0.61$\pm$0.10  & 3 \\
$\xi$ & 2.19$\pm$0.10 km\,s$^{-1}$ & 3 \\
$[$M/H$]$ & $-2.10$ & 3 \\
$d$ ({\em Gaia}) & 4.6$^{+0.7}_{-0.5}$ kpc & 4 \\
$\mu_{\alpha}$ & $-$6.66$\pm$0.05 & 1 \\
$\mu_{\delta}$ & \phs0.50$\pm$0.04 & 1 \\
\hline                  
\end{tabular}
\tablebib{(1): \citet{Gaia2018Lindegren}; (2): \citet{GaiaDR2}; 
(3): This work; 
(4): \citet{BailerJones2018}.}
\end{table}

\section{Chemical abundance analysis}
We performed a standard abundance analysis that employed a mixture of equivalent width (EW) measurements, 
carried out via Gaussian fits with the IRAF {\em splot} task, and 
spectrum synthesis. Here we employed the same line list as in 
\citet[][see Table~2]{Koch5897} with further additions  in the syntheses from 
\citet{Biemont2000},
\citet{DenHartog2003},
\citet{DenHartog2006},
\citet{Lawler2007},
\citet{Lawler2008},
\citet{Lawler2009},
\citet{Sneden2009}, and 
\citet{CJHansen2013}.

Hyperfine splitting was included where appropriate.
\begin{table}[htb]
\caption{Linelist}
\centering
\begin{tabular}{ccccc}
\hline\hline       
$\lambda$ &  & E.P. &  &  EW \\
$[$\AA$]$ & \rb{Species} & [eV] &\rb{log\,$gf$} & [m\AA] \\
\hline                  
   6300.31 &     O\,{\sc i} &       0.00  &    $-$9.819  &      \phantom{1}42 \\
   6363.79 &    O\,{\sc i} &       0.02  &    $-$10.303  &      \phantom{11}7 \\
   5682.63 &    Na\,{\sc i} &      2.10 &  $-$0.700   &     \phantom{1}46 \\
   5688.20 &    Na\,{\sc i} &     2.10  &   $-$0.460  &     \phantom{1}66 \\
   6154.23 &    Na\,{\sc i} &     2.10  & $-$1.560   &      \phantom{11}9 \\
   6160.75 &    Na\,{\sc i} &     2.10  & $-$1.260   &     \phantom{1}17 \\
   4702.99 &    Mg\,{\sc i} &     4.35  &   $-$0.440   &          141 \\
   5528.42 &    Mg\,{\sc i} &     4.35 &           $-$0.481   &     156 \\
   5711.09 &    Mg\,{\sc i} &     4.33 &           $-$1.728   &     \phantom{1}62 \\
\hline                  
\end{tabular}
\tablefoot{Table~2 is available in its entirety in electronic form via the CDS.}
\end{table}
The main abundance analysis was carried out using the ATLAS grid of one-dimensional, 72-layer, plane-parallel, line-blanketed Kurucz models without convective overshoot
and the $\alpha$-enhanced opacity distribution functions AODFNEW \citep{CastelliKurucz2003}. We further assumed that local thermodynamic equilibrium (LTE) holds for all species. 
All computations relied on the stellar abundance code MOOG \citep[][2014 version]{Sneden1973} unless noted otherwise. 
\subsection{Stellar parameters}
To derive stellar parameters, we used photometry provided by 
{\em Gaia} DR2. We populated the parameter space using computed ATLAS9 model atmosphere grids by \citet{CastelliKurucz2003}. This contains theoretical values of $G_{BP}-
G_{RP}$, $A_{G}$, E($G_{BP}-G_{RP}$), and bolometric corrections 
for each set of effective temperatures ($T_{\rm eff}$), surface gravities (log\,$g$), and metallicities ([Fe/H]) in the range 
of 3500$\leq$$T_{\rm eff}$$\leq$6000 K, 0$\leq$log\,$g$$\leq$4, and $-4\leq$[Fe/H]$\leq+0.5$. 
The reddening E($G_{BP}-G_{RP}$) was computed using the reddening law of \cite{Fitzpatrick2019}. 
In order to determine the best suite of the stellar parameters $T_{\rm eff}$ \ and log\,$g$ for our target star J110842,  
the following iterative procedure was used: 
\begin{enumerate}
\item The initial metallicity was estimated via the literature value of $-1.6$ \citep{Kunder2017} 
and $T_{\rm eff}$ was derived by interpolating in $G_{BP}-G_{RP}$;
\item The bolometric correction was derived by interpolation from the new $T_{\rm eff}$; 
\item log\,$g$ was derived using the above bolometric correction and $T_{\rm eff}$; 
\item $A_{G}$ and E($G_{BP}-G_{BP}$) were derived by interpolating in $T_{\rm eff}$; 
\item $G$ and $G_{BP}-G_{RP}$ were de-reddened using the reddening maps by \citet[][A$_V$=0.81]{Schlafly2011};
\item The procedure was iterated until the difference in temperature between successive runs
was less than 50 K.
\end{enumerate}

The microturbulence velocities ($\xi$) in each step were estimated using the calibration of \cite{Mashonkina2017}. 
Here we note that the final value of 2.19 km\,s$^{-1}$ provides an excellent balance in the plot of line-by-line abundances with equivalent widths of the 
Fe\,{\sc i} lines.

The final photometric parameter set of  $T_{\rm eff}$=4421 K and log\,$g$=0.61 dex yields an iron abundance from the neutral and ionized species
of [Fe\,{\sc i}/H]=$-1.84$ and [Fe\,{\sc ii}/H]=$-2.10$ dex, respectively. 
Thus, there is a pronounced ionization imbalance seen in this star when employing the photometrically derived surface gravity. 
Moreover, no equilibrium of the Fe\,{\sc i} abundance with excitation potential could be reached upon using the photometric temperature. 
This is a well-known problem for stars more metal poor than about $-$1.5 dex, as has been systematically evaluated by \citet{Mucciarelli2020}. 
The suggested reasons for these discrepancies are   the commonly used assumptions in the spectroscopic approach, to wit, LTE and/or the one-dimensional treatment of the atmospheres. 
Therefore, following the recommendation of \citet{Mucciarelli2020}, we adopt in the following the stellar parameters derived photometrically above and we continue by 
choosing the Fe abundance from the ionized species as the metallicity scale of star J110842.
\subsection{Abundance errors}
The statistical errors on our abundance ratios were determined via the standard deviation and the number of
measured lines per element used to derive its abundance. Furthermore, we performed a 
systematic error analysis by varying each stellar parameter about its respective uncertainty:
$T_{\rm eff}$$\pm50$ K, log\,$g\pm$0.1 dex, [M/H]$\pm$0.1 dex, and  $\xi\pm$0.1 km\,s$^{-1}$. 
We further ran the identical analyses as above using solar-scaled opacity distributions (ODFNEW) 
and take one-quarter of the ensuing deviation to 
mimic an ignorance of the $\alpha$-enhancement in the star of 0.1 dex. The respective deviations of the abundance
ratios from the {bona fide} results from the unaltered atmospheres are listed in Table~3; a conservative upper limit to the total systematic uncertainty in terms of the squared sum of all contributions
 is given in the last column, although strong correlations between the impacts from the various atmospheric parameters 
 can be expected \citep[see, e.g.,][]{McWilliam1995,Hanke2020HD20}. 

\begin{table*}[htb]
\caption{Systematic error analysis.}
\centering
\begin{tabular}{ccccccc}
\hline\hline       
 & $T_{\rm eff}$ & log\,$g$& [M/H] &  $\xi$ &  &  \\
\rb{Species} & $\pm$50 K & $\pm$0.1 dex & $\pm$0.1 dex &  $\pm$0.1 km\,s$^{-1}$ & \rb{ODF} & \rb{Sys.} \\
\hline       
CH (G-band) & $\pm$0.03  & $\mp$0.03& $\mp$0.07 & $\mp$0.01& $-$0.08 & 0.08 \\
O\,{\sc i}  &  $\pm$0.02 &  $\pm$0.04 &  $\pm$0.03 &  $\mp$0.01 &  $-$0.08 &  0.07 \\
Na\,{\sc i} &  $\pm$0.05 &  $\mp$0.02 &  $\mp$0.01 &  $\mp$0.01 & \phs0.04 &  0.06 \\
Mg\,{\sc i} &  $\pm$0.06 &  $\mp$0.03 &  $\mp$0.02 &  $\mp$0.04 & \phs0.05 &  0.08 \\
Si\,{\sc i} &  $\pm$0.02 &  $\pm$0.01 &  $\mp$0.01 &  $\mp$0.01 & \phs0.01 &  0.03 \\
Ca\,{\sc i} &  $\pm$0.07 &  $\mp$0.02 &  $\mp$0.02 &  $\mp$0.04 & \phs0.05 &  0.09 \\
Sc\,{\sc ii} &  $\mp$0.01 &  $\pm$0.03 &  $\pm$0.02 &  $\mp$0.03 &  $-$0.06 &  0.06 \\
Ti\,{\sc i} &  $\pm$0.14 &  $\mp$0.02 &  $\mp$0.03 &  $\mp$0.06 &  $-$0.03 &  0.16 \\
Ti\,{\sc ii} &  $\mp$0.01 &  $\pm$0.03 &  $\pm$0.02 &  $\mp$0.04 &  $-$0.05 &  0.06 \\
V\,{\sc i} &  $\pm$0.11 &  $\mp$0.02 &  $\mp$0.02 &  $\mp$0.01 & \phs0.03 &  0.12 \\
Cr\,{\sc i} &  $\pm$0.12 &  $\mp$0.02 &  $\mp$0.03 &  $\mp$0.05 & \phs0.02 &  0.14 \\
Mn\,{\sc i} &  $\pm$0.11 &  $\mp$0.03 &  $\mp$0.03 &  $\mp$0.03 & \phs0.05 &  0.12 \\
Fe\,{\sc i} &  $\pm$0.08 &  $\mp$0.02 &  $\mp$0.02 &  $\mp$0.05 & \phs0.04 &  0.10 \\
Fe\,{\sc ii} &  $\mp$0.04 &  $\pm$0.04 &  $\pm$0.02 &  $\mp$0.02 &  $-$0.06 &  0.07 \\
Co\,{\sc i} &  $\pm$0.09 &  $\mp$0.02 &  $\mp$0.02 &  $\mp$0.04 & \phs0.05 &  0.11 \\
Ni\,{\sc i} &  $\pm$0.06 &  $\mp$0.01 &  $\mp$0.01 &  $\mp$0.02 & \phs0.04 &  0.09 \\
Cu\,{\sc i} &  $\pm$0.08 &  $\mp$0.01 &  $\mp$0.01 &  $\mp$0.02 & \phs0.04 &  0.09 \\
Zn\,{\sc i} &  $\mp$0.03 &  $\pm$0.01 &  $\pm$0.01 &  $\mp$0.03 &  $-$0.01 &  0.05 \\
Sr\,{\sc ii} &  $\pm$0.01 &  $\pm$0.01 &  $\pm$0.02 &  $\mp$0.01 &  $-$0.05 &  0.03  \\
Y\,{\sc ii} &  $\mp$0.01 &  $\pm$0.03 &  $\pm$0.02 &  $\mp$0.04 &  $-$0.05 &  0.06 \\
Zr\,{\sc ii} &  $\mp$0.01 &  $\pm$0.03 &  $\pm$0.02 &  $\mp$0.01 &  $-$0.05 &  0.05 \\
Ba\,{\sc ii} &  $\pm$0.03 &  $\pm$0.04 &  $\pm$0.02 &  $\mp$0.08 &  $-$0.09 &  0.11 \\
La\,{\sc ii} &  $\pm$0.02 &  $\pm$0.02 &  $\pm$0.01 &  $\mp$0.04 &  $-$0.03 &  0.05 \\
Ce\,{\sc ii} &  $\pm$0.01 &  $\pm$0.03 &  $\pm$0.02 &  $\mp$0.01 &  $-$0.06 &  0.05 \\
Pr\,{\sc ii} &  $\pm$0.02 &  $\pm$0.03 &  $\pm$0.02 &  $\mp$0.01 &  $-$0.07 &  0.06 \\
Nd\,{\sc ii} &  $\pm$0.01 &  $\pm$0.02 &  $\pm$0.01 &  $\mp$0.03 &  $-$0.03 &  0.04 \\
Sm\,{\sc ii} &  $\pm$0.02 &  $\pm$0.03 &  $\pm$0.02 &  $\mp$0.01 &  $-$0.05 &  0.05 \\
Eu\,{\sc ii} &  $\pm$0.02 &  $\pm$0.02 &  $\pm$0.02 &  $\mp$0.07 &  $-$0.06 &  0.08\\
Gd\,{\sc ii} &  $\pm$0.01 &  $\pm$0.02 &  $\pm$0.01 &  $\mp$0.01 &  $-$0.04 &  0.03 \\
Dy\,{\sc ii} &  $\pm$0.02 &  $\pm$0.02 &  $\pm$0.01 &  $\mp$0.02 &  $-$0.04 &  0.04 \\
Hf\,{\sc ii} &   $\pm$0.05 &  $\pm$0.05 &  $\pm$0.05 &  $\mp$0.03 &  $-$0.02 &  0.09 \\
Pb\,{\sc ii} & $\pm$0.10 &  $\mp$0.02 &  $\mp$0.07 &  $\mp$0.01 &  \phs0.10 & 0.13  \\
\hline                  
\end{tabular}
\end{table*}

\section{Results}
All abundance results and the errors as described above are listed in Table~4. These values adopt the solar abundance scale of \citet{Asplund2009}.
In the following figures we place our results into context with the MW halo, bulge, and disks, and 
\wcenp 
For $\omega$Cen we used the data of \citet{Johnson2010} and \citet{Simpson2020}, who chemodynamically extracted cluster candidates from the GALAH survey \citep{DeSilva2015}.
Here  we also show the abundance ratios of those stars.
Figures 1, 3, and 4  show the abundance comparison, where we restrict ourselves
to those elements in common between our study and that of \wcenp 
The remaining elements, though not
explicitly shown, are discussed individually below.
\begin{table*}[htb]
\caption{
Abundance results. Abundance ratios for ionized species are given relative to Fe\,{\sc ii}. 
For iron itself, [Fe/H] is listed. The line-to-line scatter, $\sigma$, and number of measured lines, $N$, indicate the statistical 
error.}
\centering
\begin{tabular}{cccr|cccr|cccr}
\hline\hline       
Species & [X/Fe] & $\sigma$ & $N$ & 
Species & [X/Fe] & $\sigma$ & $N$ & 
Species & [X/Fe] & $\sigma$ & $N$  \\
\hline       
 CH (G-band) &   $-$0.85 & \ldots &  1$^{\rm S}$ & Mn\,{\sc i}  &  $-$0.50 &   0.12 &  9$^{\rm H}$ & La\,{\sc ii} &  \phs0.40 &   0.09 & 6$^{\rm H}$ \\
 O\,{\sc i}  &  \phs0.65 & 0.28 &  2$^{\rm S}$ & Fe\,{\sc i}  &  $-$1.91 &   0.13 & 78      & Ce\,{\sc ii} &  \phs0.13 &   0.07 & 3$^{\rm S}$ \\
Na\,{\sc i}  &  \phs0.51 &   0.05 &  4           & Fe\,{\sc ii} &  $-$2.01 &   0.05 &  8      & Pr\,{\sc ii} &  \phs0.32 &   0.07 & 5$^{\rm S}$ \\
Mg\,{\sc i}  &  \phs0.54 &   0.14 &  3           & Co\,{\sc i}  & \phs0.15 &   0.15 &  5$^{\rm H}$ & Nd\,{\sc ii} &  \phs0.19 &   0.09 & 5$^{\rm S}$ \\
Si\,{\sc i}  &  \phs0.48 &   0.06 &  2           & Ni\,{\sc i}  &  $-$0.03 &   0.10 & 15      & Sm\,{\sc ii} &  \phs0.24 &   0.07 & 3$^{\rm S}$ \\
Ca\,{\sc i}  &  \phs0.31 &   0.20 & 15           & Cu\,{\sc i}  &  $-$0.33 &   0.10 &  2      & Eu\,{\sc ii} &  \phs0.00 & \ldots & 1$^{\rm S}$ \\ 
Sc\,{\sc ii} &  \phs0.11 &   0.14 &  2$^{\rm H}$ & Zn\,{\sc i}  & \phs0.13 &   0.08 &  2      & Gd\,{\sc ii} &  \phs0.15 & \ldots & 1$^{\rm S}$ \\ 
Ti\,{\sc i}  &  \phs0.29 &   0.12 & 11           & Sr\,{\sc ii} &       $<$0.30   & \ldots & 1$^{\rm S}$ & Dy\,{\sc ii} &  \phs0.23 &   0.13 & 3$^{\rm S}$ \\
Ti\,{\sc ii} &  \phs0.38 &   0.15 & 10           &  Y\,{\sc ii} & \phs0.10 &   0.08 &  5      & Er\,{\sc ii} &   $-$0.02 & \ldots & 1$^{\rm S}$ \\ 
 V\,{\sc i}  &  \phs0.13 &   0.03 &  6           & Zr\,{\sc ii} & \phs0.36 & \ldots &  1      & Hf\,{\sc ii} &   $-$0.05 & \ldots & 1$^{\rm S}$ \\ 
Cr\,{\sc i}  &   $-$0.06 &   0.07 &  7           & Ba\,{\sc ii} & \phs0.23 &   0.01 &  3$^{\rm H}$ & Pb\,{\sc i}  &  \phs0.60 & \ldots & 1$^{\rm S}$ \\ 
\hline                  
\end{tabular}
\tablefoot{``H'' indicates that hyperfine structure was accounted for; ``S'' denotes abundances that were derived from spectrum synthesis.}
\end{table*}
\subsection{Metallicity}
At [Fe\,{\sc ii}/H]=$-$2.10 dex, J110842 samples the metal-poor tail of $\omega$Cen's metallicity distribution (Fig.~1, top).
This distribution has long been known to show a large dispersion and covers a range of more than 1.7 dex \citep{Johnson2010}.
As this object is commonly considered the nucleus of a formerly more massive dwarf galaxy,  
such a large spread and the occurrence of very metal-poor stars down to $-2.5$ dex \citep{Johnson2020wcen} is not surprising as this is seen 
in many dwarf spheroidal galaxies in the Local Group \citep[e.g.,][]{Koch2009Review}.
\begin{figure}[htb!]
\begin{center}
\includegraphics[angle=0,width=1\hsize]{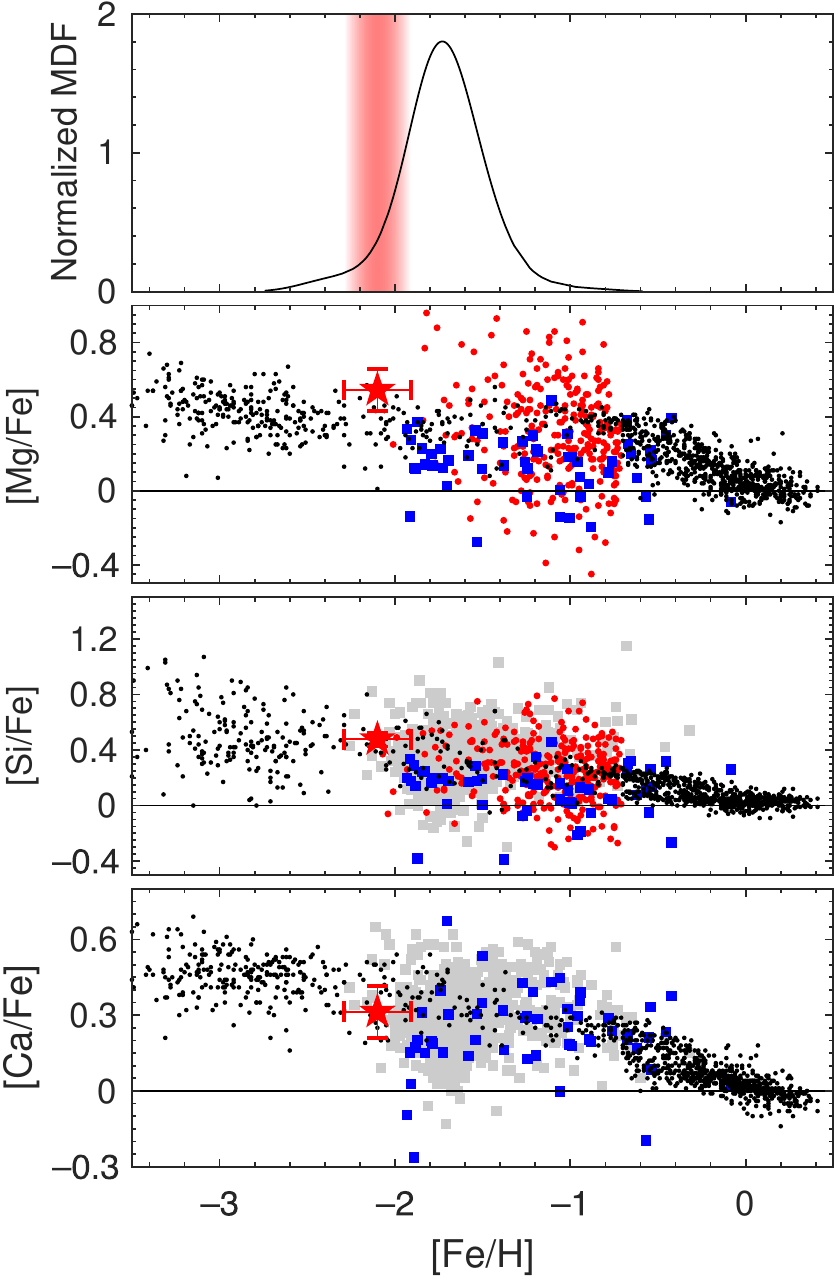}
\end{center}
\caption{Chemical abundances of J110842 (red star) in comparison with \wcen (gray squares: \citealt{Johnson2010}; blue squares: \citealt{Simpson2020}) and the MW halo \citep{Roederer2014}
and disks \citep{Bensby2014}, shown as black dots.  The 329 candidate  $\omega$Cen associates from \citet{FernandezTrincado2015} are indicated  
as red points.
The   error bar accounts for statistical and systematic uncertainties.
 The top panel shows the error-weighted metallicity distribution from 
\citet{Johnson2020wcen}, on which is highlighted the 1$\sigma$ error range of the metallicity determined in the present work.}
\end{figure}
\subsection{Light elements: C, N, O, Na}
\citet{FernandezTrincado2015} noted that \wcen is special in that it covers a broad range in all of their analyzed abundance patterns
 to the point that it overlaps with all MW components. They concluded that any similarity in these
properties is ``not very useful'' to constrain the origin of the stars. 
However, their element abundances, drawn from the RAVE survey, were only limited to Fe, Al, and Ni\footnote{Abundances for the other $\alpha$-elements
(Mg, Si, Ti) were reported for the remainder of the $\omega$Cen candidates in \citet{FernandezTrincado2015}, but had not been derived for star J110842.}.
Therefore, we highlight here some of the elements that provide a greater clue to  any potential origin of J110842.

One of the  characteristics of GCs is the presence of multiple stellar generations and the ensuing light-element variations
as a result of $p$-capture reactions in an early generation of stars \citep{Kayser2008,Carretta2009NaO,Bastian2018}. 
These variations include bimodalities in CN and associated anticorrelations with CH strength.
We measured the strength of the commonly used 
CH and CN bands using the index definitions of \citet{Norris1981}, which was then translated into a $\delta$S(3839) index to remove dependencies on
evolutionary status \citep{Martell2010,Koch2019CN}.
The main uncertainty in this quantity is the required absolute magnitude of the star, which relies on its distance that still shows large 
errors (see Table~1). 
At $\delta$S(3839)=0.253, J110842 qualifies as a CN-strong star (see, e.g., Fig.~5 in \citealt{Koch2019CN}). As such it would be tempting to 
characterize it as a second-generation star, bolstering its origin in a GC-like environment, which also requires it to be CH-weak.  
We measured an S(CH) index of 0.819, and combined with the lower [C/Fe] we conclude that this instead argues in favor of  a C-normal star
 \citep[e.g.,][]{Kirby2015,Koch2019Pal15}, 
with values that are appropriate for its luminosity of 1224 $L_{\odot}$ \citep{GaiaDR2}.

Similarly, genuine GCs are infamous for having a pronounced Na-O anticorrelation, which is also prominently seen in \wcen 
\citep{Norris1995NaO,Gratton2011,Simpson2020} and particularly extended below $-$1.3~dex \citep{Marino2011}.
In Fig.~2 we show our Na and O measurements for J110842 superimposed on the literature data from the above sources. 
\begin{figure}[htb!]
\begin{center}
\includegraphics[angle=0,width=1\hsize]{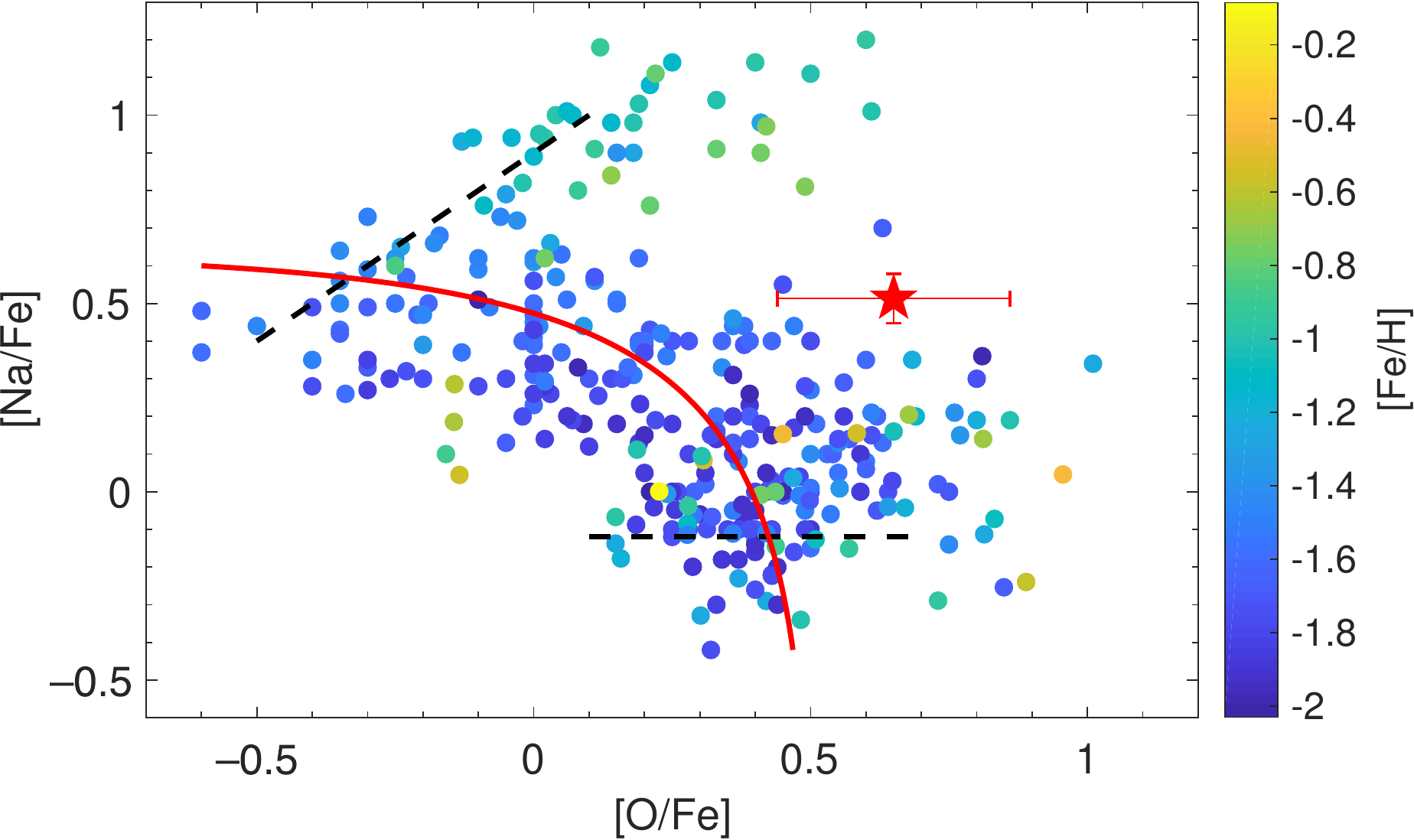}
\end{center}
\caption{Sodium-oxygen anticorrelation using data from \citet{Marino2011} and \citet{Simpson2020}, color-coded by metallicity. Star J110842 is indicated by a red star.
We also indicate the empirical separations into first, second, and extreme generations (dashed black) and the simplistic dilution model (red line) by \citet{Carretta2009NaO}.}
\end{figure}

While the value of  [O/Fe]  for J110842 at { 0.65 dex} is  compatible with that of a metal-poor, $\alpha$-enhanced halo star, the [Na/Fe] value of 0.55 dex is rather high
and makes this star fall into the regime of Na-strong second-generation GC stars. We note that all our abundances have been derived in an LTE framework.
However, interpolating the grid of non-LTE (NLTE) corrections by \citet{Lind2011} yields a departure from the LTE abundance of $\sim-0.06$ dex, while 
the correction for oxygen is null for the parameters similar to our star \citep{Sitnova2013}. Thus, even if NLTE corrections are accounted for, we cannot exclude a 
second-generation origin for this star based on its {Na abundance, and marginally supported by its O abundance and CN strength}.

Finally, we note that GCs often show strong variations in Al that mildly correlate with Mg, owing to the hot branches of proton-burning.
However, the spectral range of our UVES setting did not allow us to determine an Al abundance
from the 6696~\AA\  line. Conversely, the blue line at 3961~\AA~lies in the wing of the strong Ca H line, making a meaningful abundance determination difficult.
Instead, we   use the value of \citet{FernandezTrincado2015}  for further discussion.   
Considering the higher Mg and O abundances in our star, their adopted [Al/Fe] of 0.38 dex lies  at the low branch of Al abundances, which is consistent with 
an association with a second stellar population in a GC \citep[e.g.,][]{Carretta2013Al}. 
\subsection{$\alpha$-elements: Mg, Si, Ca, Ti}
The $\alpha$-elements in J110842 present  few surprises. At an [$\alpha$/Fe] value of 0.42$\pm$0.03 dex it falls square on the $\alpha$-plateau delineated by 
metal-poor halo stars and the bulk of $\omega$Cen's broad abundance space (Figs.~1,3). 
This indicates enrichment via standard nucleosynthesis in supernovae of type II (SNe~II) and does not allow us to 
further investigate the question of a peculiar origin of this star based on these chemical tracers. 
\subsection{Fe-peak elements: Sc, V, Cr, Mn, Co, Ni, Cu}
As is true for the $\alpha$-elements,  the Fe-peak elements also follow the trends outlined by metal-poor halo and $\omega$Cen stars (see Figures~3,4, and also 
\citealt{Cohen1981,Norris1995abun,Smith1995,Pancino2011,Magurno2019} for $\omega$Cen)
and that are mainly set by SN Ia nucleosynthesis \citep[e.g.,][]{Kobayashi2006}.
\begin{figure}[htb!]
\begin{center}
\includegraphics[angle=0,width=1\hsize]{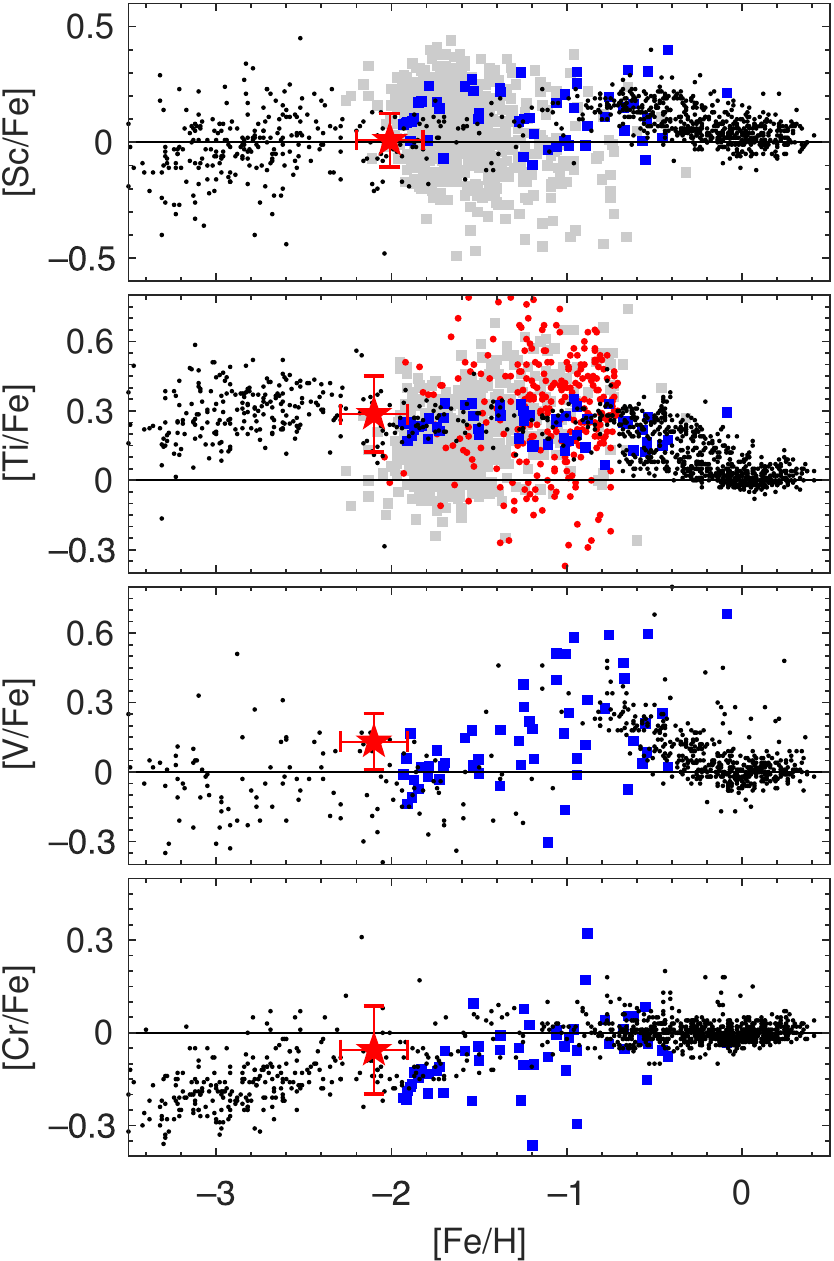}
\end{center}
\caption{Same as Fig.~1, but for the lighter Fe-peak elements. Disk values for Sc and V are   from \citet{Battistini2015}.}
\end{figure}

Copper (bottom panel of Fig.~4), in contrast, has been a matter of high interest in this GC, and 
\citet{Cunha2002} noted that Cu stays approximately constant and remains below the trend seen in halo stars over a broad
metallicity range of $\sim -2$ to $-0.8$ dex. This was thought as being due to a lower-level contribution of SNe Ia to the chemical 
evolution in that metallicity range. 
Furthermore, by chemical similarity to the Sagittarius dwarf galaxy, 
\citet{McWilliam2005} lend support to the notion that \wcen is rather the nucleus of a former, more massive system.
The target of this study, J110842, overlaps with the metal-poor halo and the metal-poor tail of the $\omega$Cen distribution.
Our own values and the cited literature values were derived under the
assumption of LTE. It appears that the Cu abundances are affected
by NLTE  effects \citep{Andrievsky2018,Shi2018}, although different
model-atoms and NLTE codes provide different results, especially
at low metallicity. This notion is reinforced by the observed
deviation between \ion{Cu}{i} and \ion{Cu}{ii}
abundances observed by \citet{RoedererBarklem2018}.
\citet{Bonifacio2010} found  strong granulation effects
on the resonance lines (although not used here); however, the combined
effect of granulation and NLTE effects still needs to
be investigated. It would therefore be interesting
to reinvestigate the Cu abundances in $\omega$Cen, and
in J110842, using a more sophisticated modeling.
\begin{figure}[htb!]
\begin{center}
\includegraphics[angle=0,width=1\hsize]{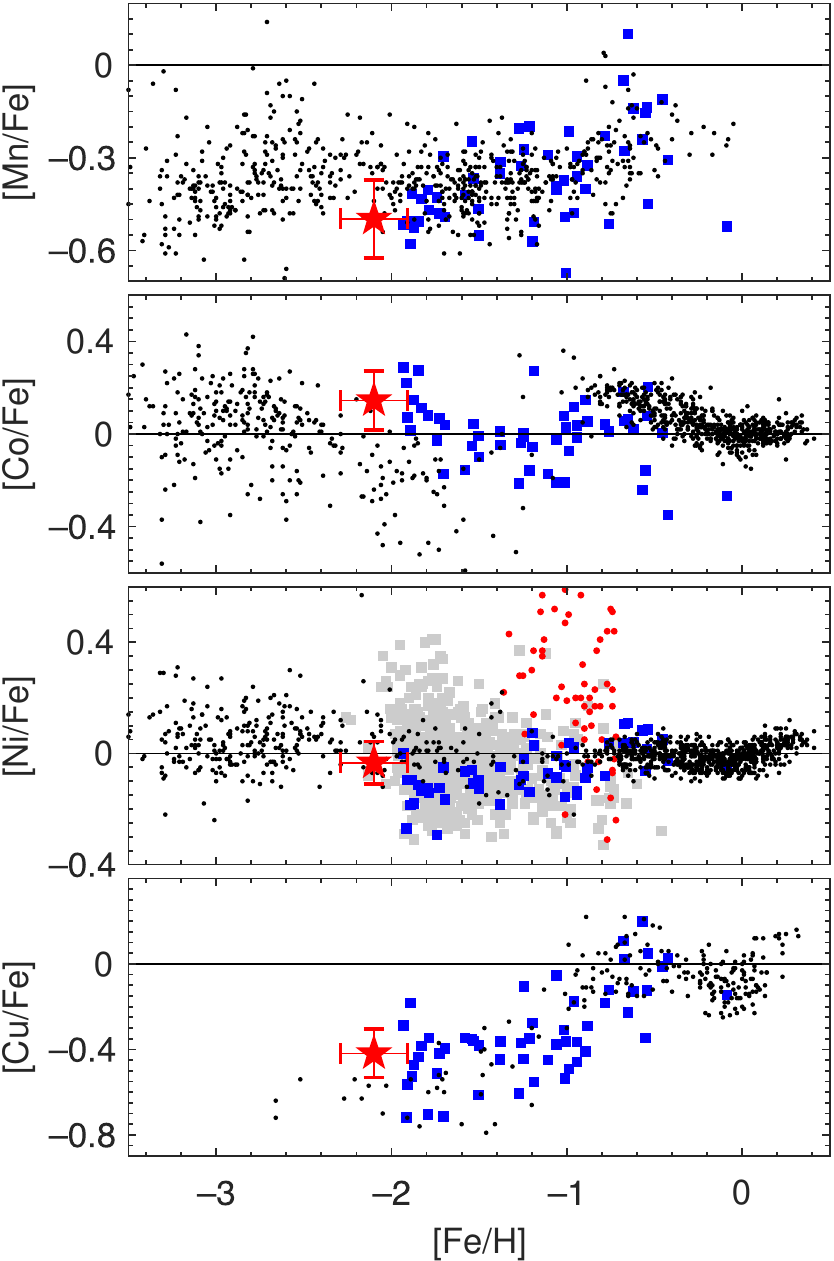}
\end{center}
\caption{Same as Fig.~1, but for the remaining Fe-peak elements. Abundances for Mn are  from \citet{Sobeck2006}, Co data are from  \citet{Battistini2015}, and 
Cu for the MW disks and halo are from \citet{Mishenina2002,Mishenina2011}.}
\end{figure}

We can also use our measurements to address the  broader context of Galactic chemical evolution. Here 
  \citet{Hawkins2015} posited that the [Mg/Mn] versus [Al/Fe] plane is a powerful indicator for an  origin in major, dwarf-galaxy-like accretion events versus {in situ} formed stars that 
are enhanced in the $\alpha$-elements. 
In J110842 we measured a very high value for  [Mg/Mn]  of 1.04$\pm$0.09 dex. 
If we had taken into account NLTE effects,
this overabundance would be even larger \citep{Bergemann2019}.
In order to qualify as an accreted object, the [Al/Fe] abundance in J110842 would need to be subsolar at its Mg and Mn abundances, according to 
the distinctive line in \citet{Horta2020}. However, since \citet{FernandezTrincado2015} reported on a higher [Al/Fe] of 0.38 dex, 
this instead argues 
for an {in situ} formation or that the birth environment has chemically evolved.
\subsection{$n$-capture elements: Zn, Sr, Y, Zr, Ba, La, Ce, Pr, Nd, Sm, Eu, Gd, Dy, Er, Hf, Pb}
Our abundance results for the selected heavy elements overlapping with the literature for \wcen are shown in Fig.~5.
In addition,  the values for J110842 for these elements are in close agreement with  metal-poor halo stars and the metal-poor end of the GC
abundance distribution, indicating that the same nucleosynthetic processes were at play.
\begin{figure}[htb!]
\begin{center}
\includegraphics[angle=0,width=1\hsize]{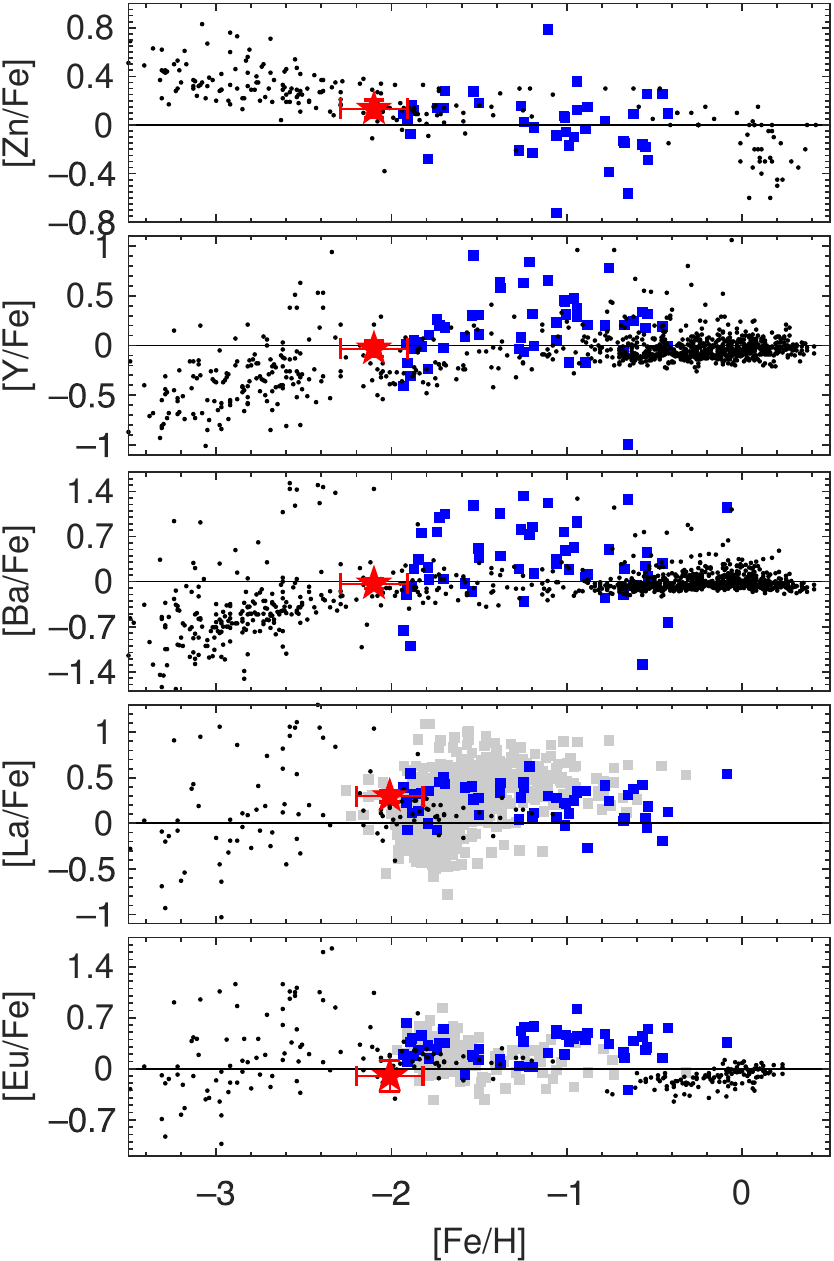}
\end{center}
\caption{Same as Fig.~1, but for the neutron-capture elements ($Z\ge$30). The reference MW abundances are  from \citet{Barbuy2015} for Zn,
and from \citet{Koch2002} for Eu.}
\end{figure}

Among the neutron-capture elements we particularly note the $s$-process elements Y, Ba, and La (middle three panels of Fig.~5).
Dating back to \citet{LloydEvans1983}, a bimodal behavior has now   been established; for instance,
from their high-resolution analysis of 113 RR Lyr stars in \wcen \citet{Magurno2019} reported 
 solar [Y/Fe] values for stars more metal poor than approximately $-$1.5 dex, while the more metal-rich component 
showed enhanced element abundances. This is also seen in terms of a significantly larger scatter above that metallicity cut
 in the data of  \citet{Simpson2020}, among others. Furthermore,   the values for [La/Fe]  found by \citet{Johnson2010} also show a  
clear bimodality. 
This has been interpreted in terms of markedly different stellar populations, where the more metal-rich 
component was (self-)enriched over a long timescale by low-mass asymptotic giant branch stars 
\citep[e.g.,][]{Norris1995abun,Smith2000,Cunha2002}, 
while the sudden increase in the abundance ratios with [Fe/H] is  compatible with \wcen being the remnant of 
a more massive dwarf galaxy \citep{Romano2007,Magurno2019}. 
The solar value of [Y/Fe] in J110842 at its low metallicity and   its solar [Ba/Fe] are fully compatible with the lower-metallicity component.

Figure~6 shows the overall abundance distribution of heavy ($Z$$>$30) elements we were able to measure in J110842. 
Here we note that Sr in its spectrum shows very strong resonance lines (at 4077 and 4215~\AA) that are likely saturated. 
Therefore, we  were only able to place a limit of $\sim$0.3 dex on the [Sr/Fe] value. 
\begin{figure}[tb!]
\begin{center}
\includegraphics[angle=0,width=1\hsize]{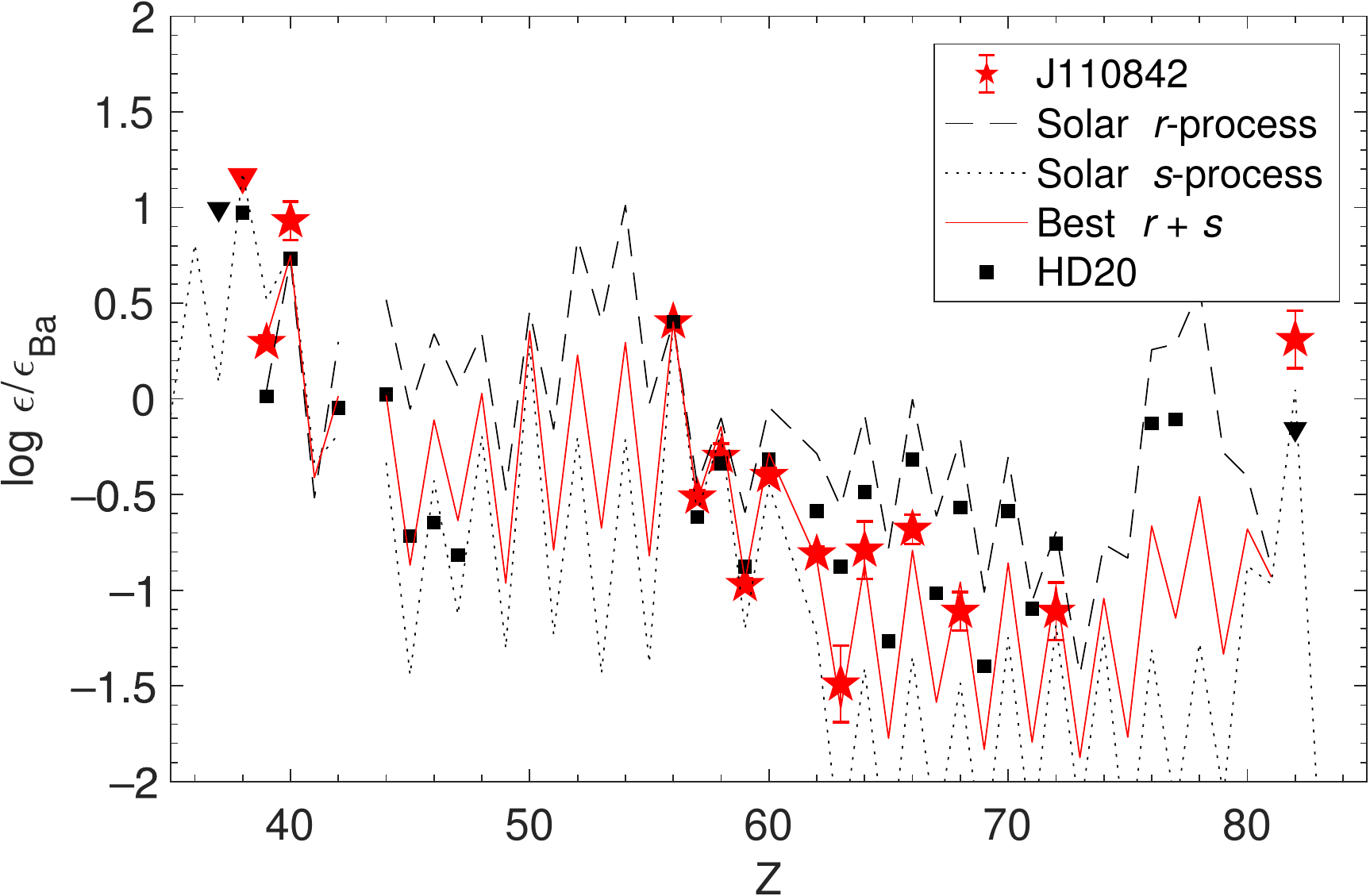}
\end{center}
\caption{Abundance distribution of the heavy elements, normalized to Ba. Also shown are the solar $s$- and $r$-process contributions \citep{Burris2000} and the best-fit
linear combination of the contributions.
Finally, we  overplot the $r$-process benchmark star HD\,20 as black squares \citep{Hanke2020HD20}.}
\end{figure}
As noted before, J110842 is mainly characterized by standard nucleosynthesis and the majority of elements lie between the (solar) $s$- and $r$-process
distributions, as is expected if we consider,  at this metallicity, that AGB stars have already started to contribute some $s$-process material 
to the Galactic chemical evolution \citep[e.g.,][]{Simmerer2004}.
A $\chi^2$ fit indicates that J110842 has received an admixture of $s$-process elements at the 60\% level.
 Another assessment of the $r$- and $s$-process contributions can be made in comparison to the cleaner $r$-process tracer HD\,20 (black symbols in Fig.~6), 
 which at a  metallicity that is lower than the Sun's displays a very clean $r$-process pattern.
 However,  in this direct comparison J110842 also comes out as a $r$--$s$ mixed 
 star with not just one main polluter governing its formation.
A further characterization of the donors to the chemical enrichment of this star, such as SNe or AGB stars, 
cannot be unambiguously made in stars like J110842, where the effects of mixing and dilution need to be properly dealt with \citep{Magg2020,CJHansen2020}. 
\section{Discussion}
In our endeavor to find signs of a chemical association of the halo star J110842 with the massive GC $\omega$Cen, as
has been previously suggested from their relative dynamics, we   performed an extensive chemical abundance study.
We showed that the majority of the heavier elements ($Z$$\ge$12) is fully compatible with those seen in \wcenp 
This is, however, not surprising given the large overlap of the GC abundance space with that of metal-poor halo field stars at this metallicity ($-2.1$ dex), 
and cannot unambiguously argue for a GC origin of J110842.
Here it is more interesting to look at the light elements. The high Na abundance and its characterization as a CN-strong star indeed lend support to the 
hypothesis that it is a former second-generation GC star, while the enhancement in O and its CH-normal nature appear to mitigate this. 

\citet{FernandezTrincado2015} argued that it is unlikely that stars with high relative encounter speeds have been ejected from $\omega$Cen, but 
just had chance encounters when their orbits coincided. 
The central escape velocity of the GC is on the order of 60--100 km\,s$^{-1}$ \citep{Gnedin2002,Gao2015}, which  
compares to the computed encounter velocity with J110842 in excess of 200 km\,s$^{-1}$. 
On the other hand, extreme horizontal branch 
stars at high tangential motions of up to 90--310 km\,s$^{-1}$ have been found in $\omega$Cen, and they are most likely to escape the GC within the 
next Myr or so \citep{Gao2015}, in which case a former origin of our target star within the GC remains plausible. 
Here we note that \citet{Lind2015} associate a halo star with typical GC signatures such as enhanced Al and low Mg abundances with having escaped from \wcen at high speed.
Such high-speed ejections ($>$100 km\,s$^{-1}$) are suggested if interactions with binaries or black holes are
evoked \citep{Gvaramadze2009,Luetzgendorf2012}.

As to the origin of \wcen itself, \citet{Myeong2019} have associated it with the recently discovered massive 
Sequoia accretion event, while \citet{Ibata2019} paired it with the Fimbulthul stream.
The latter poses an intriguing parallel to our star as its abundance distribution is very similar to the most metal-poor stream candidate
analyzed by \citet{Simpson2020}. This emphasizes the power of chemical tagging in 
meaningfully investigating the manifold of eclectic GC-stream-halo-field interfaces. 
\begin{acknowledgements}
The authors thank the anonymous referee for a swift and constructive report.
AJKH gratefully acknowledges funding by the Deutsche Forschungsgemeinschaft (DFG, 
German Research Foundation) -- Project-ID 138713538 -- SFB 881 (``The Milky Way System''), subprojects A03, A05, A11. 
CJH acknowledges support from the Max Planck Society and from the ChETEC COST Action (CA16117), supported by COST (European Cooperation in Science and Technology).
LL, PB and EC gratefully acknowledge support
from the French National Research Agency (ANR) funded project
``Pristine'' (ANR-18-CE31-0017).
This work has made use of data from the European Space Agency (ESA) mission
{\it Gaia} (\url{https://www.cosmos.esa.int/gaia}), processed by the {\it Gaia}
Data Processing and Analysis Consortium (DPAC,
\url{https://www.cosmos.esa.int/web/gaia/dpac/consortium}). Funding for the DPAC
has been provided by national institutions, in particular the institutions
participating in the {\it Gaia} Multilateral Agreement.
This research made use of atomic data from the INSPECT database, version 1.0 (\url{www.inspect-stars.com}).
 \end{acknowledgements}
\bibliographystyle{aa} 
\bibliography{ms} 
\end{document}